\input epsf.tex  

\documentclass[preprint2]{aastex}

\newcommand{\simless}{\mathbin{\lower 3pt\hbox
     {$\rlap{\raise 5pt\hbox{$\char'074$}}\mathchar"7218$}}} 
\newcommand{\simgreat}{\mathbin{\lower 3pt\hbox
     {$\rlap{\raise 5pt\hbox{$\char'076$}}\mathchar"7218$}}} 
\slugcomment{Submitted 20 July 1999 ; Revised 26 Aug 1999 }

\shorttitle{Koresko et al.}
\shortauthors{Haro 6--10 Companion}

\begin{document}


\title{ Imaging the Haro 6--10 Infrared Companion }


\author{ Chris D. Koresko, Geoffrey A. Blake, Michael E. Brown}
\affil{ Division of Geological and Planetary Sciences, Caltech, Pasadena, CA 91125 }
\author{ Anneila I. Sargent }
\affil{ Department of Astronomy, Caltech, Pasadena, CA 91125 }
\author{ David W. Koerner }
\affil{ Department of Physics \& Astronomy, U. Pennsylvania, Philadelphia, PA 19104-6396 }

    
\begin{abstract}

We present an infrared imaging study of the low-mass pre-main sequence
binary system Haro 6--10.  This system is one of a handful in which the
optically-visible primary has the characteristics of a normal T~Tauri
star, while the secondary is a so-called ``infrared companion" (IRC), a
strongly extincted object which emits most of its luminosity in the
infrared.  A speckle holographic technique was used to produce nearly
diffraction-limited images on three nights over a one-year period starting
in late 1997.  The images show that the IRC is obscured and surrounded by
a compact, irregular, and variable nebula.  This structure is in striking
contrast to the well-ordered edge-on disk associated with HK~Tauri~B, the
extincted companion to another T~Tauri star of similar age.  A new,
resolved intensity peak was found 0\arcsec.4 southwest of the IRC.  We
suggest that it may represent light scattered by a clump of dusty material
illuminated by starlight escaping along an outflow-carved cavity in the
IRC envelope.  The primary star became fainter and the companion became
more extended during the observing period.

\end{abstract}


\section{Introduction}


Among the low-mass pre-main sequence binary systems in nearby active
star-forming regions, there exist a handful in which one of the two stars
is reminiscent of a protostar, radiating primarily at infrared wavelengths
and faint or undetected in visible light.  These objects are referred to
as ``infrared companions" (IRCs), despite being in most cases more
luminous than their primaries.  Their bolometric temperatures, which
measure the ``center of mass" of the spectral energy distribution and are
correlated with evolutionary status for young stars (Myers \& Ladd 1993),
tend to lie in a transition region between true embedded sources on the
one hand and classical or weak-lined T~Tauri stars on the other.  A
variety of models have been proposed for the IRCs, but they are generally
taken to be dust-shrouded stars which are either less evolved than their
primaries and have yet to dissipate their natal envelopes ({\it e.g.,}
Dyck, Simon, \& Zuckerman 1982) or are experiencing episodes of enhanced
accretion, perhaps due to interactions with the primary or with a
circumbinary disk ({\it e.g.,} Koresko, Herbst, \& Leinert 1997; hereafter
KHL).  The prototype IRC is the companion to T~Tauri itself.

Early studies of Haro 6--10 by Elias (1978) revealed a nonstellar
photographic appearance, a spectrum displaying prominent forbidden line
emission, and photometric variations of a factor $\sim 3$ at 2.2 $\mu$m. 
The infrared companion was found 1\arcsec.2 north of the visible star by
Leinert \& Haas (1989; hereafter LH), who used slit-scanning speckle
interferometry to measure the ratio of the brightnesses of the two stars
at wavelengths between 1.65 and 4.8 $\mu$m.  As in the T~Tauri system, the
IRC was fainter than the visible star at wavelengths shorter than $\sim 4$
$\mu$m, but brighter in the mid-infrared.  The Haro~6--10~IRC is the
reddest object studied by KHL, with a bolometric temperature of only
210~K, compared to 490~K for the T~Tauri IRC.

In addition to the IRC, LH found evidence of {\it extended} emission, in
the form of visibility amplitude curves which fell below unity at high
frequencies, for scans taken perpendicular to the axis of the binary.  By
contrast, scans taken along the binary axis showed only the oscillations
typical of a pair of pointlike stars.  LH argued that this extended
emission was probably associated with one of the two stars.  With the
advent of  two-dimensional infrared detectors on large telescopes,
together with refined speckle imaging techniques, it has become practical
to directly image the components of young binaries to search for faint
tertiary components and diffuse circumstellar material.  This {\it Letter}
presents the results of a speckle holographic imaging study of the Haro
6--10 IRC.

\section{Observations}

The new high-resolution images were taken in the K (2.2 $\mu$m)  photometric
band at the 10~m Keck~1 telescope on three nights in November 1997, March
1998, and November 1998 (see Table 1) using the Near-Infrared Camera
(Matthews \& Soifer 1994).  The NIRC Image Converter (Matthews et al. 
1996) produced a magnified pixel spacing of 0\arcsec.02, approximately
Nyquist-sampling the diffraction limit at 2 $\mu$m.  The observations
consisted of thousands of exposures of the visual binary, with integration
times of $\sim 0.15$ sec for each frame.  This short exposure time
partially ``froze" the atmospheric seeing, so that the point-spread
function (PSF) consisted of distinguishable speckles.  The 1\arcsec.2
separation of the binary was large enough that the PSFs did not
significantly overlap in frames with good seeing.

Individual frames were calibrated in the standard way by subtracting mean
sky frames, dividing by flatfield images, and ``fixing" bad pixels.  A
model was computed for the ``bleed" signal which extended along and
perpendicular to the readout direction, and this was subtracted from the
calibrated frame.  For each frame, a measurement of the instantaneous
point-spread function (PSF) was made by masking all pixels on the side of
the frame toward the IRC, leaving only the primary.   Frames in which the
instantaneous seeing was too poor for this procedure to cleanly separate
the stars were rejected.  For the rest, the Fourier power spectrum of the
PSF frames, and the cross-spectrum ({\it i.e.,} the Fourier transform of
the cross-correlation) of the masked frame with the unmasked frame, were
computed.  If the primary star is unresolved, then in principle the ratio
of the cross-spectrum to the PSF frame's power spectrum is the Fourier
transform of the diffraction-limited image.

Raw frames produced by NIRC suffer from semi-coherent electronic pattern
noise which is typically concentrated in a small regions of the Fourier
domain.  If not corrected, this noise limits the sensitivity of the
holography technique.  To identify the contaminated frequencies, a noise
frame was constructed by replacing the pixels containing the stars 
with copies of an empty region near a corner of the field.  The power
spectrum of the noise frame was computed, and frequencies containing more
power in the noise frame than the average over the series of unmasked
frames at similar frequencies were marked as bad.  

The PSF power and the cross-spectrum at uncontaminated frequencies were
accumulated over the whole series of frames.  A final image was
reconstructed from them, with the use of an apodizing function to suppress
high-frequency noise.  The image was then rotated to standard
orientation.  The apodizing function chosen was the product of a Gaussian
and a Hanning function.  It produced a final image resolution of 81~mas
(FWHM).  

The resulting images for each of the three nights are presented in
Figure~1.  The image from each epoch is scaled to the maximum pixel level in
the IRC.  These images display a dynamic range, measured as the brightness
of the peak of the primary star in units of noise, of $\simgreat 2000$. 
This is much larger than is typical for a speckle image; the improvement is
due to the holography technique's use of instantaneous PSF measurements
instead of relying on a statistical consistency of the atmosphere between
separate target and reference-star frames as in normal speckle
interferometry.  The technique does produce a residual artifact in the form
of an apparent ridge of emission which runs approximately midway between the
two stars and parallel to the direction of the mask used to make the PSF
frames.  This is probably due to a small amount of flux from the primary
which spills over into the masked region.  In addition, in the final epoch
there is a narrow strip which extends from the IRC in a direction
approximately north and exactly parallel to the detector readout.

A direct K-band image of the region was taken without the image converter,
so that the raw field was 38\arcsec~($\sim 5000$ AU).  Frames were taken
with the binary in two well-separated locations, and the reduction was
done by computing the difference between these frames and pixel-fixing. 
The resulting image, presented in Figure~2, has much lower resolution but
higher flux sensitivity than the holographic image.  It shows the two
stars surrounded by an arc of nebulosity which curves gently from the
primary star.

\begin{deluxetable}{crrrrr}
\footnotesize
\tablecaption{ Holographic Observations
\label{Observations}}
\tablewidth{0pt}
\tablehead{
\colhead{Date (UT)} & \colhead{$\lambda$ ($\mu$m)} & \colhead{$\tau_{int}$} & 
\colhead{Frames} & \colhead{Seeing FWHM} & \colhead{Brightness Ratio}
}
\startdata
22 Nov 1997 & 2.2 $\mu$m & 0.15 & 931 & 0\arcsec.32 & $0.153 \pm .001$ \\
8 Mar 1998 & 2.2 $\mu$m & 0.15  & 1276 & 0\arcsec.36 & $0.295 \pm .001$ \\
3 Nov 1998 & 2.2 $\mu$m & 0.14 & 2230 & 0\arcsec.40 & $0.388 \pm .002 $ \\
\enddata
\end{deluxetable}

\section{ Results }

The Haro~6--10 infrared companion is clearly resolved in the image taken
on each of the three nights.  Its peak is marginally resolved in all three
images, with a deconvolved FWHM of $\sim 35$ mas, or 5 AU, along cuts in
the East-West direction.  The peak is surrounded by a complex and
irregular structure, the brightest part of which consists of a nebulous
``tail" that extends some 300 mas to the south.  A fainter peak, which we
designate Haro~6--10~IRC--SW, appears 400 mas to the southwest of the
IRC.  Finally, a narrow ``arm" extends 500 mas westward from a point just
north of the peak.

The IRC underwent a significant morphological change during the
observations, especially between the second and third epochs.  Between the
first and third epochs, a new ``tail" appears to the north of the peak,
reaching an extent of $\sim 200$ mas.  It is fainter but wider than the
southern tail.  At the same time, IRC--SW dimmed by a factor of
$\sim 2$ compared to the IRC's peak. 

The 2.2 $\mu$m brightness of the IRC, measured as a fraction of the
brightness of the primary, more than doubled from 0.15 to 0.39 (see
Table~1).  For comparison, LH found it to be 0.13 in September 1988, and KHL
found 0.04 in October 1994.  

In order to compare the holographic images with the one-dimensional (1-D)
speckle measurements by LH, we projected the holographic images along
position angle perpendicular to 355$^\circ$, computed the 1-D Fourier
amplitudes, and divided them by the amplitude of a projected point-source
image made with the same apodization function.  The first and third epochs
show excellent agreement with the K-band amplitude curve of LH, while the
middle epoch shows small ($\sim 5\%$) depression in the zero-frequency
level.

\begin{figure*}[htbp]
\epsfxsize=15.0cm
\centerline{
  \epsfbox{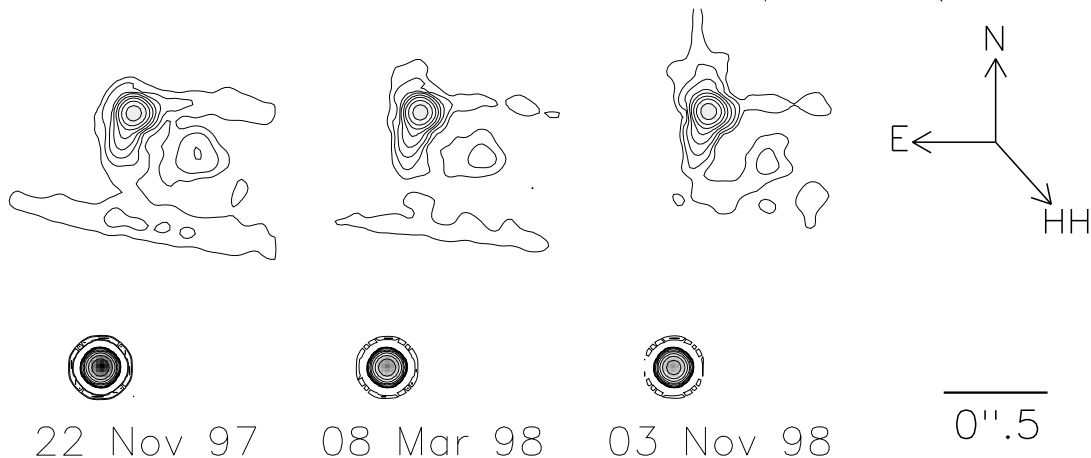}
}
\caption[ ]{ 
Holographic images of the Haro~6--10 system at 2.2 $\mu$m taken on three
nights over a 1-year period.  They are scaled so that the peak flux in the
IRC is the same for the three epochs.  The stars are separated by
1\arcsec.2, and the resolution in the images is 81 mas (FWHM).  The images
are in standard orientation with north up, east left.  The IRC's peak is
resolved, and it is surrounded by a complex distribution of circumstellar
matter.   A bright ``tail" extends to the south and, in the final epoch, a
fainter one extends to the north.  The nebulous peak IRC--SW lies
0\arcsec.4 to the southwest of the IRC, along the direction of the giant
Herbig-Haro flow (labeled ``HH" in the compass rose). The primary is
unresolved by assumption, and the rings around its image are intrinsic to
the PSF produced by the post-processing.  The ridgelike structure between
the stars and the narrow stripe extending north from the IRC in the final
epoch are artifacts. 
}
\end{figure*}

\begin{figure*}[htbp]
\epsfxsize=10.0cm
\centerline{
  \epsfbox{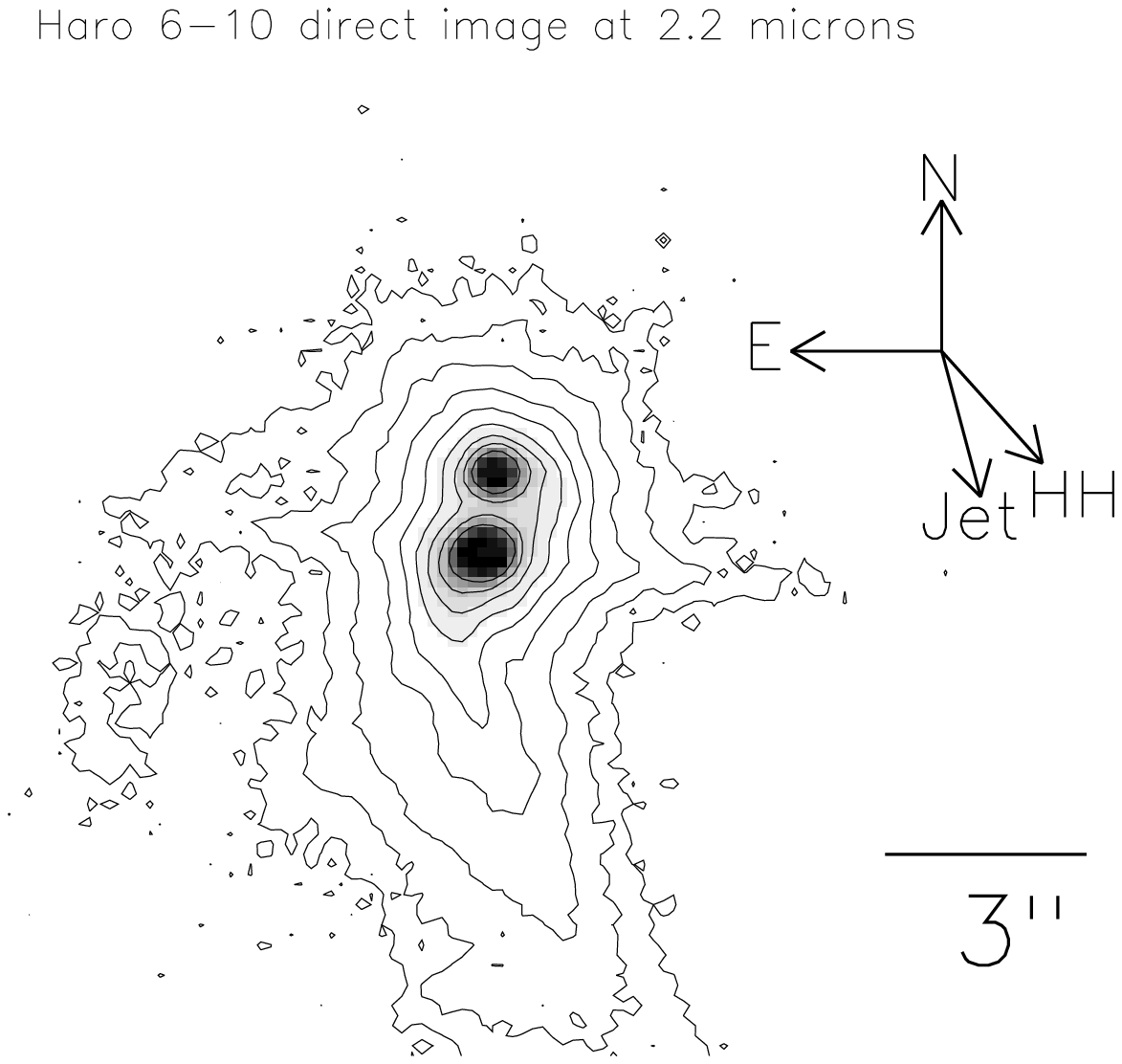}
}
\caption[ ]{ 
This direct image of Haro~6--10 and the IRC was made by stacking four
1.0-second K-band frames taken without the image converter.  It shows the
two stars and the small jet previously detected by Movessian \& Makagian
(1999), which is seen to curve gently away from the primary star.  Note
that the faint spikes in this image are due mainly to diffraction from the
hexagonal secondary spider and the edges of the segments which make up the
primary mirror.  The image is in standard orientation, and the contour
levels are peak $\times 2^{-(1...9)}$.
}
\end{figure*}

\section{ Discussion }

It would be natural to suppose that the circumstellar material surrounding
the IRC may be in the form of an optically-thick disk of gas and dust. 
Its ``peak with tails" morphology in the final epoch qualitatively
resembles simulations of a nearly edge-on disk ({\it e.g.,} Wood et al.
1998), with the southern and northern tails tracing light scattering in
the upper regions of the Sunward-facing side of the disk, and the star
located somewhere slightly to the west of the emission peak.  Such a disk
is seen in HK~Tauri, in which the secondary star is completely obscured
and the disk is traced at visible and near-infrared wavelengths via the
starlight it scatters (Stapelfeldt et al. 1998; Koresko 1998).  In the
case of Haro~6--10, the disk axis would be nearly perpendicular to the
line joining the two stars, as one would expect if the disk lies in the
orbital plane of the binary, and in contrast to the probable geometry of
the HK~Tauri system.  

The large extinction which would be produced by an edge-on disk is
consistent with previous infrared spectroscopic and spectrophotometric
results.  A deep mid-infrared silicate absorption feature was seen toward
the IRC by van Cleve et al. (1994), while no such absorption was seen
toward the primary.   The low-resolution K-band spectrum measured by
Herbst, Koresko, \& Leinert (1995) is a featureless continuum except for a
molecular hydrogen $v = 1-0$ S(1) emission line at 2.12 $\mu$m.  This
suggests that the 2 $\mu$m light originates primarily in dust, which may
be heated either by stellar photons or by gravitational energy released
via disk accretion.  The lack of a detectable $v = 2-1$ S(1) line was
taken to indicate that the hydrogen is excited in a shock, presumably
associated with either accretion or outflow, rather than pumped by
ultraviolet photons.  

However, the simple edge-on disk picture alone fails to account for
several important features of the system.  In particular, it is not
obviously consistent with the existence of Haro~6--10~IRC--SW and the
west-facing arm, or with the lack of a northern tail in the first epochs. 
If most of the circumstellar mass does reside in a disk, then it appears
that the disk may have suffered strong perturbations which have disrupted
its outer regions.

The timescale over which the IRC's morphology changes is too short to
correspond to orbital or freefall motions in the material in the outer
regions of the nebula, where the orbital period is on the order of
centuries.  This suggests that the changes are the result of changing
illumination of the distant material due, {\it e.g.,} to shadowing of the
central star by material orbiting within $\sim 1$ AU of the star, or by
starspots ({\it e.g.,} Wood \& Whitney 1998).

Additional clues to the structure of the IRC are provided by the outflows
associated with the system.  A giant Herbig-Haro flow which extends about
1.6 pc (39 arcmin) along a position angle close to 222$^\circ$ was found
by Devine et al. 1999.  These authors suggest that the flow originates
from the IRC, which may be problematic in the context of the disk model,
as the outflow's position angle is $\sim 45$$^\circ$ from the
sky-projected axis of the putative disk.  Movessian \& Makagian (1999)
report the discovery of a jet which curves away from the binary at a
position angle of 195$^\circ$, although they could not identify its source
with either of the stars.  This is presumably structure seen in our direct
image, in which it appears to be associated with the primary.  

If the IRC is the source of the giant H-H flow, then Haro~6--10~IRC--SW
lies suggestively along the flow axis.  But this object shows no sign of
the $\sim 300$ mas outward motion which would be expected over the $\sim
1$ yr span of the observations if it moved at the $\sim 200$ km~s$^{-1}$
typical of the pattern speed in such a flow ({\it e.g.,} Eisloffel \&
Mundt 1992).  By contrast, the free-fall velocity 
would produce an undectable motion.
We conclude that IRC--SW is more likely to be orbiting the IRC or
associated with envelope material than to be part of the outflow.

If IRC--SW were a {\it stellar} companion to the IRC, it could easily
perturb the IRC's disk, and any disk of its own would be similarly
perturbed, perhaps accounting for its photometric variability.  A triple
system would likely be unstable to significant orbital evolution or even
ejection on timescales shorter than the age of Haro~6--10 (Pendleton \&
Black 1983).  On the other hand, IRC--SW's surface brightness is roughly
consistent with a model in which an isotropically-scattering dusty surface
is illuminated by unextincted light from the central star of the IRC,
which is taken to have T$\sim 5400$ K and L$\sim 6~{\rm L}_\odot$ (KHL). 
The low extinction seen by IRC--SW could result from its position along
the outflow axis of the giant H-H flow, which would clear a path through
the IRC envelope, and the scattering surface would be either an
irregularity in the surface of the outflow cavity or a knot of material
within it.  Instabilities in the outflow (Devine et al. 1999) might
account for the apparent variability of the object.






%
%
%
%
%

\section{ Conclusions }

The complex structure seen in the Haro~6--10~IRC is in stark contrast to
the simple, beautifully regular shape of the nearly edge-on disk which
surrounds HK~Tauri~B.  While it would be premature to draw strong
conclusions from this about the IRC class as a whole, it does suggest that
at least in this object the characteristic low infrared color temperature
may be the product of reprocessing and scattering in a disk which has been
strongly perturbed, at least in its outer regions.  The origin of this
perturbation is not clear, but possibilities include interactions with the
giant Herbig-Haro flow, residual infalling cloud material, or with a
possible star embedded in IRC--SW.

\acknowledgments

It is a pleasure to thank A. Bouchez for his assistance with the
observations, F. Roddier for a useful suggestion regarding the data
reduction, and the anonymous referee for suggestions which improved the
interpretation of the results.  Data presented herein were obtained at the
W.M.  Keck Observatory, which is operated as a scientific partnership
among the California Institute of Technology, the University of California
and the National Aeronautics and Space Administration.  The Observatory
was made possible by the generous financial support of the W.M.  Keck
Foundation.  This research was supported by the National Aeronautics and
Space Administration.

\clearpage

\clearpage

\end{document}